\documentclass{iaus}
\usepackage{graphicx}

\title[EAGLE Spectroscopy Beyond the Local Group] 
{EAGLE Spectroscopy of Resolved Stellar Populations Beyond the Local Group}

\author[Evans et al.]   
{Chris Evans$^1$, Yanbin Yang$^2$, Mathieu Puech$^2$, Matthew Lehnert$^2$, Michael Barker$^3$, 
Annette Ferguson$^3$, Jean-Gabriel Cuby$^4$, Simon Morris$^5$, G\'{e}rard Rousset$^6$, Fran\c{c}ois Ass\'{e}mat$^6$, 
Hector Flores$^2$}
\affiliation{$^1$UK ATC, Royal Observatory Edinburgh, 
Blackford Hill, Edinburgh, EH9 3HJ, UK \\ email@ {\tt chris.evans@stfc.ac.uk}\\
$^2$GEPI, Observatoire de Paris, 5 Place Jules Janssen, 92195 Meudon Cedex, France\\
$^3$IfA, Royal Observatory Edinburgh, Blackford Hill, Edinburgh, EH9 3HJ, UK\\
$^4$LAM, OAMP, 38 rue Fr\'{e}d\'{e}ric Joliot Curie, 13388 Marseille Cedex 13, France\\
$^5$Department of Physics, Durham University, South Road, Durham, DH1 3LE, UK\\
$^6$LESIA, Observatoire de Paris, 5 Place Jules Janssen, 92195 Meudon Cedex, France
}

\pubyear{2009}
\volume{262}  
\pagerange{1-4}
\jname{Stellar Populations}
\editors{G.~Bruzal \& S.~Charlot eds.}
\begin{document}

\maketitle

\begin{abstract}
We give an overview of the science case for spectroscopy of resolved
stellar populations beyond the Local Group with the European Extremely
Large Telescope (E-ELT). In particular, we present science simulations
undertaken as part of the EAGLE Phase~A design study for a
multi--integral-field-unit, near-infrared spectrograph.  EAGLE will exploit
the unprecedented primary aperture of the E-ELT to deliver AO-corrected
spectroscopy across a large (38.5 arcmin$^2$) field, truly revolutionising our
view of stellar populations in the Local Volume.

\end{abstract}

\firstsection 

\section{Introduction}\label{intro}

As we look ahead to the end of the coming decade, we will be entering
the era of the Extremely Large Telescopes (ELTs).  Plans are well
advanced for three ELT projects with filled apertures well in excess
of current optical-IR facilities: the Giant Magellan Telescope (GMT,
with an equivalent diameter of 21.4m), the Thirty Meter Telescope (TMT)
and the 42m European ELT (E-ELT).  When coupled with sophisticated
adaptive optics (AO) systems to correct for atmospheric turbulence,
the ELTs will provide us with unique views of stellar populations in
the Local Volume.  Although planning a decade ahead may seem far
removed, the lead-in times on these ambitious projects is more akin to
those commonly associated with space missions, i.e. construction
planning, detailed science simulations, conceptual instrument designs
and the financial planning are all well advanced on all three ELT
projects.  Here we introduce the EAGLE Phase A instrument study for the E-ELT, 
highlighting its performance for observations of resolved stellar populations.

\vspace{-0.15in}
\section{Resolved Stellar Populations: Today \& in the ELT Era}

Photometric methods are immensely powerful when applied to
extragalactic stellar populations, but only via precise chemical
abundances and stellar kinematics can we break the age-metallicity
degeneracy, while also disentangling the populations associated with
different structures, i.e. follow-up spectroscopy is required.  Over
the past decade the Calcium Triplet (CaT, spanning 0.85-0.87 $\mu$m)
has become an increasingly used diagnostic of stellar metallicities
and radial velocities in nearby galaxies, providing new views of their
star-formation histories and sub-structure, e.g. the VLT-FLAMES DART
large programme \cite{t04}.  However, 8-10m class telescopes are
already at their limits in pursuit of CaT spectra of the evolved
populations in galaxies at distances greater than $\sim$300~kpc,
e.g. Keck-DEIMOS observations in M31 struggled to yield useful
signal-to-noise below the tip of the red giant branch at $I$~$>$~21.5 \cite{c06}.

With its vast primary aperture and excellent angular resolution, the
E-ELT will be {\em the} facility to unlock spectroscopy of evolved
stellar populations in the broad range of galaxies in the Local
Volume, from the edge of the Local Group, out towards the Virgo
Cluster. This will bring a wealth of new and exciting target galaxies
within our grasp, spanning a broader range of galaxy morphologies,
star-formation histories and metallicities than those available to us
at present in the Local Group.  These observations can then be used to
confront theoretical models to provide a unique view of galaxy
assembly and evolution.  There are many compelling and ground-breaking
targets for stellar spectroscopy with the E-ELT including, in order of distance:
\smallskip
\begin{itemize}
\item{NGC 3109 and Sextans A with sub-SMC metallicities ($Z$$<$0.2$Z_\odot$), both at 1.3 Mpc.}
\item{The spiral dominated Sculptor `Group' at 2-4 Mpc.}
\item{The M83/NGC5128 (Centaurus A) grouping at $\sim$4-5 Mpc.}
\item{NGC 3379, the nearest normal elliptical at 10.8 Mpc.}
\item{The Virgo Cluster of galaxies at 16-17 Mpc, the nearest massive cluster.}
\end{itemize}
\smallskip
In contrast to proposed E-ELT observations of high-redshift
galaxies, targets for CaT spectroscopy are readily available.  For
example, deep ground-based and {\it HST} imaging in galaxies in the
Local Volume has begun to investigate their stellar populations
\cite{r05,ghosts,angst}, yet the stellar magnitudes are well beyond
spectroscopy with existing facilities.  Note that although we have
focussed mostly on southern-hemisphere targets here, there are equally compelling
northern hemisphere targets, including the M81 group, and deeper
studies in M31 and M33.

\vspace{-0.15in}
\section{EAGLE: A Multi-IFU, Near-IR Spectrograph for the E-ELT}

The EAGLE Phase A study is a French-UK partnership to provide an
advanced conceptual design of an AO-corrected, near-infrared
spectrograph with multiple integral-field units (IFUs). The baseline design
is summarised in Table~1, and has been shaped by five top-level
science topics:
\smallskip
\begin{itemize}
\item{Physics and evolution of high-redshift galaxies}
\item{Detection and characterisation of `first light' galaxies}
\item{Galaxy assembly and evolution from stellar archaeology}
\item{Star-formation, stellar clusters and the initial mass function}
\item{Co-ordinated growth of black holes and galaxies}
\end{itemize}
\smallskip
EAGLE will employ multi-object adaptive optics (MOAO, \cite{falcon} to
provide significantly improved image quality for selected target
fields within the focal plane.  This entails an array of six laser
guide stars and five natural guide stars (NGS) to map the atmospheric
turbulence.  The deformable mirror in the telescope (M4) will be used
to correct for the low-order wavefront error terms, with the
high-order terms corrected by deformable mirrors in each science
channel. An integral part of the EAGLE project is the CANARY on-sky
demonstrator on the William Herschel Telescope in La Palma
\cite{canary}.

The consortium has calculated MOAO point-spread functions (PSFs) which
take into account real NGS configurations, illustrative of relatively
good and poor performance given the spatial distribution and magnitude
of the available guide stars.  In the following section we summarise EAGLE
observations of resolved stellar populations beyond the Local Group,
which would be used to probe the assembly history and chemical evolution 
of the host galaxies.

\begin{table}[h]
\begin{center}
\caption{EAGLE Baseline Design. The patrol field is the instrument field-of-view
within which IFUs can be configured to observe individual targets.}\label{specs}       
\begin{tabular}{ll}
\noalign{\smallskip}\hline
Parameter & Specification \\
\hline
Patrol Field & Eqv. 7$^\prime$ diameter \\
IFU field-of-view & 1{\mbox{\ensuremath{.\!\!^{\prime\prime}}}}65 $\times$ 1{\mbox{\ensuremath{.\!\!^{\prime\prime}}}}65 \\
Multiplex (\# of IFUs) & 20 \\
Spatial resolution & 30\%EE in 75 mas ($H$ band)\\
Spectral resolving power ($R$) & 4,000 \& 10,000\\
Wavelength range & 0.8-2.5$\mu$m\\
\hline
\end{tabular}
\end{center}
\end{table}

\section{EAGLE Performance: CaT Spectroscopy}

Simulated EAGLE observations of the CaT region were computed using a
modified version of the IFU tool developed to characterise the MOAO
requirements for ELT observations of high-redshift galaxies
\cite{p08}.  A synthetic CaT spectrum (T$_{\rm eff}$\,=\,4,000\,K, log{\it g} = 2.0) is adopted as a template from the Kurucz
model atmosphere calculations for the {\it GAIA} mission \cite{mc00}.
The synthetic spectra were calculated for $R=$20,000,
i.e. sufficiently over-sampled so as to be degraded to either of the
two spectral resolving powers provided by EAGLE.  To this template we
appended an additional continuum region to provide a line-free region
with which to investigate to the signal-to-noise (S/N) of the spectra
in the resulting datacubes.  The principal assumptions in the
simulations were:
\smallskip
\begin{itemize}
\item{Telescope: 42-m primary, with central obscuration of 9\%}
\item{Exposure time: 10 hrs (20x1800s)}
\item{$\lambda$-range: 8400-8750 \AA, at $R$ = 10,000}
\item{Spatial sampling: 37.5 mas spatial pixels}
\item{Total throughput: 0.19 [telescope (0.8); atmos. (0.95); EAGLE (0.25 @ 0.85$\mu$m)]}
\item{Detector: read-out noise: 5e$^-$/pixel \& dark current: 0.01e$^-$/pixel/s}
\end{itemize}
\smallskip
Assuming a Paranal-like site, we have investigated two sets
of seeing conditions in the simulations:
0{\mbox{\ensuremath{.\!\!^{\prime\prime}}}}65 at $\lambda$\,=\,0.5$\mu$m at zenith
(the mean VLT seeing at Paranal; \cite{sm08}) and
0{\mbox{\ensuremath{.\!\!^{\prime\prime}}}}90 at a zenith distance
(ZD) of 35$^\circ$, providing a good investigation of the performance
from execution of a `Large Programme'-like survey.  

The signal-to-noise (S/N) recovered, as a function of $I$-band
magnitude, for the two NGS configurations is given in
Table~\ref{sims1}.  These results are for spectra extracted from the
central spatial pixel of a point source at the centre of the cube
(optimal PSF-fitting extractions should be able to improve on these
results).  The key result is that {\it S/N\,$\ge$\,10 is recovered
from a stacked 10\,hr exposure at I\,=\,24.5, in mean seeing, in both
NGS configurations}.  This corresponds to spectroscopy of stars at the
tip of the red giant branch (RGB), with M$_I$\,=\,$-$4, out to
$\sim$5\,Mpc in just 10\,hrs.  This is four magnitudes deeper than
FLAMES using the LR08 grating ($R$\,$=$\,6,500) with the same
exposure time. Similar calculations at $R$~=~4,000 for $I$\,=\,24.5 and 26.0 (with the latter
approx. the tip of the RGB in NGC\,3379) yield S/N\,$\ge$\,10 in 5 and 80\,hrs, respectively.

\vspace{-0.1in}
\begin{table}[h]
\begin{center}
\caption{EAGLE CaT results: Continuum S/N obtained for $R$ = 10,000, $t_{\rm exp}$ = 10 hrs. }\label{sims1}       
\begin{tabular}{ccccc}
\hline
& \multicolumn{2}{c}{Seeing = 0{\mbox{\ensuremath{.\!\!^{\prime\prime}}}}9 @ ZD=35$^\circ$} &
\multicolumn{2}{c}{Mean VLT Seeing (0{\mbox{\ensuremath{.\!\!^{\prime\prime}}}}65 @ ZD=0$^\circ$)} \\
$I_{\rm VEGA}$ & NGS `good' & NGS `poor' & NGS `good' & NGS `poor' \\
\hline
22.5 &  40 &  27 & 56 & 48 \\
23.5 &  16 &  11 &  28 & 24 \\
24.5 &  $\phantom{1}$8 & $\phantom{1}$4 & 13 & 10 \\
\hline
\end{tabular}
\end{center}
\end{table}

\smallskip

\begin{figure}
\begin{center}
\begin{tabular}{cc}
\resizebox{0.4\columnwidth}{!}{\includegraphics{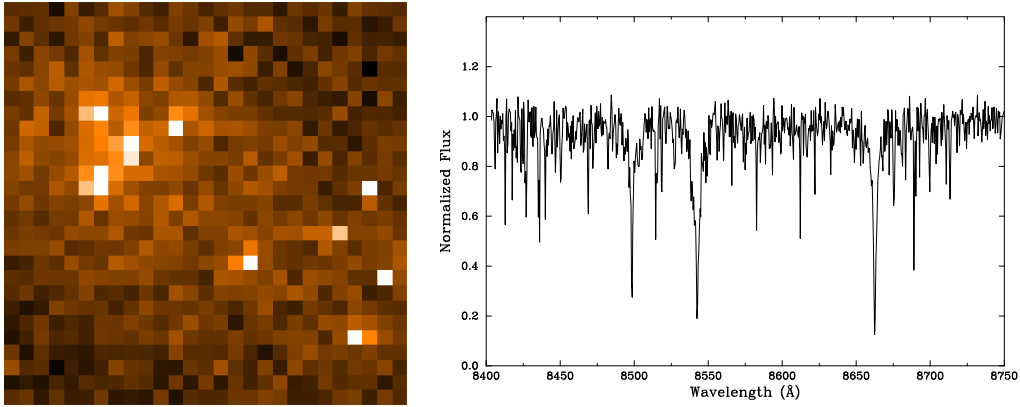}} & \resizebox{0.6\columnwidth}{!}{\includegraphics{run324.ps}}
\end{tabular}
\caption{{\it Left:} Front slice of simulated IFU datacube for EAGLE observations in the central region of NGC\,55; 
{\it Right:} Simulated CaT spectrum for a star with $I$ = 23.5, yielding S/N $\sim$ 25 in the continuum.}\label{data}
\end{center}
\end{figure}

An example EAGLE Large Programme is to undertake spectroscopy of
evolved stars in the five spiral galaxies in the Sculptor Group, by observing
multiple EAGLE fields across the major and minor axes of each galaxy.
These galaxies represent the most immediate opportunity to study the
star-formation history and mass assembly of spirals beyond the limited
sample available at present, i.e. the Milky Way, M31 and M33.
The left-hand panel of Figure~\ref{data} shows the central
1$^{\prime\prime}\,\times\,$1$^{\prime\prime}$ of an IFU observation
in the core region of NGC\,55 (at 1.9\,Mpc), with the right-hand panel
showing a simulated CaT spectrum.  
The magnitudes and relative positions of the stars are
from {\it HST} imaging in the core region of NGC\,55, taken as part of
the GHOSTS survey (\cite{ghosts}).  This example illustrates perfectly
the gain in effective multiplex from the IFUs (i.e. nine stars in this
1$^{\prime\prime}\,\times\,$1$^{\prime\prime}$ region), with minimal
impact from crowding.  

\vspace{-0.125in}
\section{Summary}
EAGLE provides a unique combination of abilities to harness the power
of the E-ELT for spectroscopy of resolved stellar populations.  The
image quality from MOAO will be significantly better than that
obtained from seeing-limited or ground-layer AO modes, enabling us to
explore spatially-resolved, extragalactic stars across a wide
field of more than five arcminutes.  A range of large programmes can
be envisaged, each of which will help provide a fundamentally new view
of stellar populations in the local Universe; from mapping of the
Sculptor galaxies, to deeper observations of the most luminous 
evolved stars in selected fields in galaxies at 10\,Mpc and beyond.

\end{document}